\documentclass[review]{elsarticle}

\usepackage{lineno,hyperref}
\modulolinenumbers[5]
\usepackage{multicol}
\usepackage[dvipsnames]{xcolor}
\journal{Journal of \LaTeX\ Templates}

\begin{document}

\begin{frontmatter}

\title{Substrate surface effects on electron-irradiated graphene}
%\tnotetext[mytitlenote]{Fully documented templates are available in the elsarticle package on \href{http://www.ctan.org/tex-archive/macros/latex/contrib/elsarticle}{CTAN}.}

%% Group authors per affiliation:
%\author{Federica Bianco\fnref{myfootnote}}
%\address{Piazza San Silvestro 12,56127, Pisa, (Italy)}
%\fntext[myfootnote]{Since 1880.}

%% or include affiliations in footnotes:
\author[CNR]{Luca Basta}
\author[SNS]{Aldo Moscardini}
\author[CNR]{Stefano Veronesi}
%\ead[url]{www.elsevier.com}

\author[CNR]{Federica Bianco\corref{mycorrespondingauthor}}
\cortext[mycorrespondingauthor]{Corresponding author}
\ead{federica.bianco@nano.cnr.it}
\ead{Tel: +39 050 509378}

\address[CNR]{NEST Laboratory, Istituto Nanoscienze-CNR and Scuola Normale Superiore, Piazza San Silvestro 12, 56127, Pisa, Italy}

\address[SNS]{NEST Laboratory, Scuola Normale Superiore, Piazza San Silvestro 12, 56127, Pisa, Italy}

\begin{abstract}
Chemical, mechanical, thermal and/or electronic properties of bulk or low-dimensional materials can be engineered by introducing structural defects to form novel functionalities. When using particles irradiation, these defects can be spatially arranged to create complex structures, like sensing circuits, where the lateral resolution of the defective areas plays a fundamental role. Here, we show that structural defects can be patterned by low-energy electrons in monolayer graphene sheets with lateral resolution strongly defined by the surface of the supporting substrate. Indeed, two-dimensional micro-Raman mapping revealed that the surroundings of irradiated areas contain unintentional defects of the graphene lattice, whose density depends on the methods exploited to clean the supporting surface. By combining Monte Carlo simulations with the analysis of the graphene Raman modes, we attributed these structural modifications mainly to the action of back-scattered electrons and back-scattered electrons-created secondaries. The latter create reactive radicals at the interface between graphene and the supporting surface that affect the lateral resolution of the defective areas. Hence, defects pattern can be produced with high lateral resolution by removing any organic contaminants from the supporting surface and by reducing the thickness of the substrate, in order to minimize the number of back-scattered and secondary electrons. 
\end{abstract}

\end{frontmatter}

\section{Introduction}
%\subsection{Graphics}
Structural defects engineering is a versatile approach for altering chemical, mechanical, thermal and/or electronic properties of bulk and low-dimensional materials to create novel properties and functionalities \cite{Jiang2019, Park2018, Jangizehi2020, Feng2019,Meynell2020}. Hence, a precise and reliable control of these defects is a fundamental paradigm for developing innovative structures with notable technological impact. 
\\At low-dimensional scale, graphene is the most studied two-dimensional material. In its pristine form, it is largely employed in optoelectronic devices \cite{Wangrev2019}, flexible electronics \cite{Shrivas2020}, heat dissipation films and energy storage \cite{Zhao2019} thanks to its excellent optical, thermal and electronic properties. On the other hand, the intentional insertion of structural defects can expand the applications of graphene in those fields that are barely reached in defect-free conditions. For example, defects modify the thermal properties of graphene via defect-phonon scattering contribution, resulting in thermal conductivity tailored by defects population \cite{Malekpour2016} and in enhancement of the otherwise poor thermoelectric performances \cite{Anno2017}. Theoretical and experimental studies have also demonstrated the relevance of the structural defects in enhancing the surface chemical reactivity of graphene sheet both in the presence (vacancy-type defects) or the absence (topological-type defects like Stone-Wales and reconstructed vacancies) of dangling bonds  \cite{Yang2018,Boukhvalov2008,Hernandez2013, Ye2017}. Hence, precise control in defects formation yields to the fine tailoring of the surface chemistry of graphene, which is fundamental when engineering its electronics properties \cite{Zhang2011, Yang2011, Nourbakhsh2010} or, more in general, for sensing applications \cite{Lee2016, Cho2016, Yuan2013, Kislenko2020}.  The most used configuration for graphene-based sensors relies on field-effect transistors (GFETs) \cite{Fu2017}. Their performances have been improved by using defect-rich graphene, as reported in radiation \cite{Yu2015}, pressure \cite{Mohammad2015}, chemical \cite{Lee2016} and gas sensors \cite{Ma2019}, as well as in biosensors \cite{Fu2017}. In such devices, the enhanced chemical reactivity necessitates also an accurate patterning onto the chip surface in order to design the optimal functionalities and/or create complex sensing circuits and large sensors array. The most versatile approach that satisfies this requirement is based on particles irradiation techniques, i.e. on the exposure of graphene sheet to focused beams of energetic particles, such as ions
or electrons (see Ref. \cite{Yang2018} and references therein). While ions or high energetic ($>$50 keV) electrons exposure involves elaborated and expensive equipment and is limited to small chip areas, effective defects modulation can be clearly patterned over a large area via electrons generated by widely diffused and cheaper scanning electron microscopes (SEMs) \cite{Childres2010,Liu2011,Childres2014,Tao2013}. Indeed, electrons with kinetic energy in the range between 5 keV and 30 keV can transform the graphene lattice from crystalline to nanocrystalline up to amorphous \cite{Tao2013, Teweldebrhan2009} and induce enough defects for chemical functionalization \cite{Malekpour2016, Lan2014} or modification of the electronic properties \cite{Liu2011}. On one hand, complex sensing circuits require a full control of the spatial distribution of the defects population, which is typically ensured by coupling pattern generator systems to SEMs. On the other hand, surroundings of irradiated areas should not be affected by unintentional modifications of the material properties. 

Although the large literature reporting the creation of structural defects via low energy electrons, the lateral resolution of defects patterning has not deeply investigated (to the best of our knowledge). In this article, we show that surface treatments of the substrate supporting graphene sheet play a key role in improving the lateral resolution of defective regions induced by low-energy electron-irradiation. Two-dimensional micro-Raman mapping revealed that unintentional defects are created in the surroundings of the exposed areas, whose extension and density strongly depend on the used surface cleaning method. Monte Carlo simulations pointed out that these defects are produced by the back-scattered electrons (BSEs) and by the interaction of the secondary electrons (SEs) generated near the substrate surface by BSEs with organic impurities (for instance hydrocarbons) that are adsorbed on silicon/silicon dioxide (Si/SiO2) substrate, as shown by the full agreement between simulation and experimental data. 

\begin{figure}[t]
\centering
  \includegraphics[height=9cm]{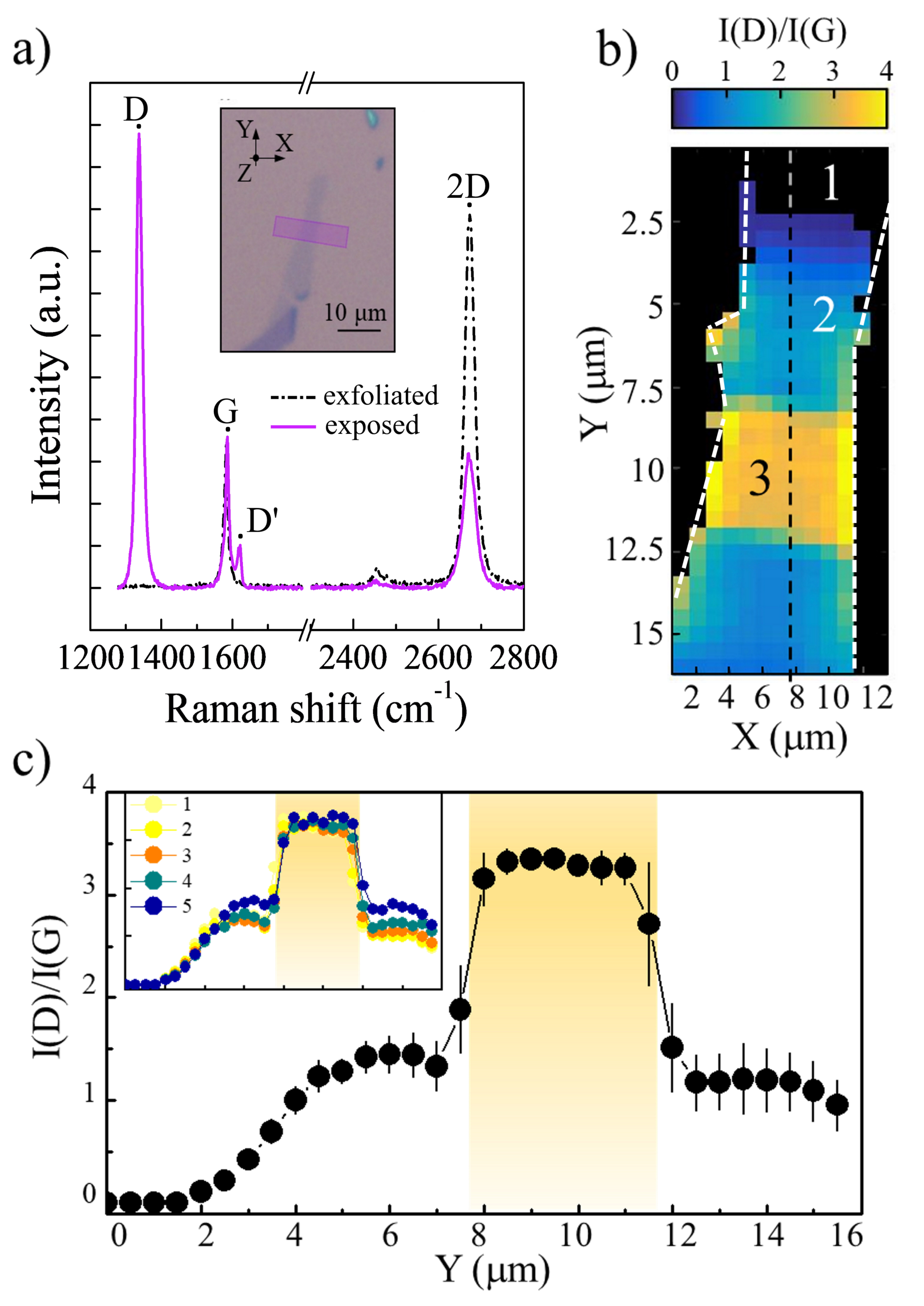}
  \caption{Defects analysis: a) representative Raman spectrum of as-exfoliated (black dash-dotted curve) and e-beam exposed graphene (violet curve) on no-plasma-treated substrate. Inset is the optical image of the graphene flake, where the violet area indicates the exposed area. The electron kinetic energy is 30 keV and dose is 40 mC/cm$^2$. b) Two-dimensional spatial distribution of I(D)/I(G). The white dashed lines indicate the flake edges. Zone indicated with 1 is the unexposed area; zone 2 is the transition area; zone 3 is the irradiated region. c) Averaged I(D)/I(G) profile along Y-axis extracted around the reference line [black dashed line in panel b)].  The error bar quantifies the variation of I(D)/I(G) among the pixels along X-axis. Inset shows 5 different profiles around the reference line [black dashed line in panel b)]. The yellow area indicates the irradiated area. The solid line is a guide for eye.}
  \label{fig1}
\end{figure}

\section{Material and methods}
Graphene flakes were micromechanically exfoliated from highly oriented pyrolytic graphite on boron-doped Si substrate having 300-nm-thick thermally-grown SiO$_2$. Metallic markers were lithographed by electron-beam (e-beam) for defining flakes positions. Before exfoliation process, the substrate surface was cleaned by e-beam resist residue removal solution (hereafter referred as "no-plasma-treated substrate") and by oxygen plasma at 100 W for 5 minutes (hereafter referred as "plasma-treated substrate"). 
\\Before e-beam exposure, graphene lattice conditions were characterized by micro-Raman spectroscopy. Flakes were scanned by laser beam at 532 nm with 100x objective, corresponding to a lateral resolution of 1 $\mu$m. A proper step-size of 0.5 $\mu$m was used to map the Raman modes across the surface.
\\A defective area was created by irradiating - in a single step - the graphene flake with electrons accelerated at 20 keV and 30 keV. This area was well-defined by a pattern generator coupled-SEM and it is 4 $\mu$m-long and as wide as the graphene flake. The dimensions of the defects-rich area were arbitrarily chosen in order to be sufficiently resolved by the used micro-Raman system. The e-beam (current of about 0.15 nA) was scanned with a step-size of 0.100 $\mu$m. The dose was varied from 5 to 200 mC/cm$^2$, resulting in a dwell-time ranging from 3 to 140 ms. 
\\The defects induced by electrons were investigated by micro-Raman spectroscopy in air-ambient just after the exposure with the same parameters of pre-exposure investigation.
\\Monte Carlo simulations were carried out by CASINO software \cite{CASINO}, considering the experimental conditions: $\sim$ 8$\times$10$^5$ and 4$\times$10$^6$  incident electrons (corresponding to about 40 mC/cm$^2$ and  200 mC/cm$^2$, respectively) and beam radius of 10 nm.

\begin{figure}[t]
\centering
  \includegraphics[height=8cm]{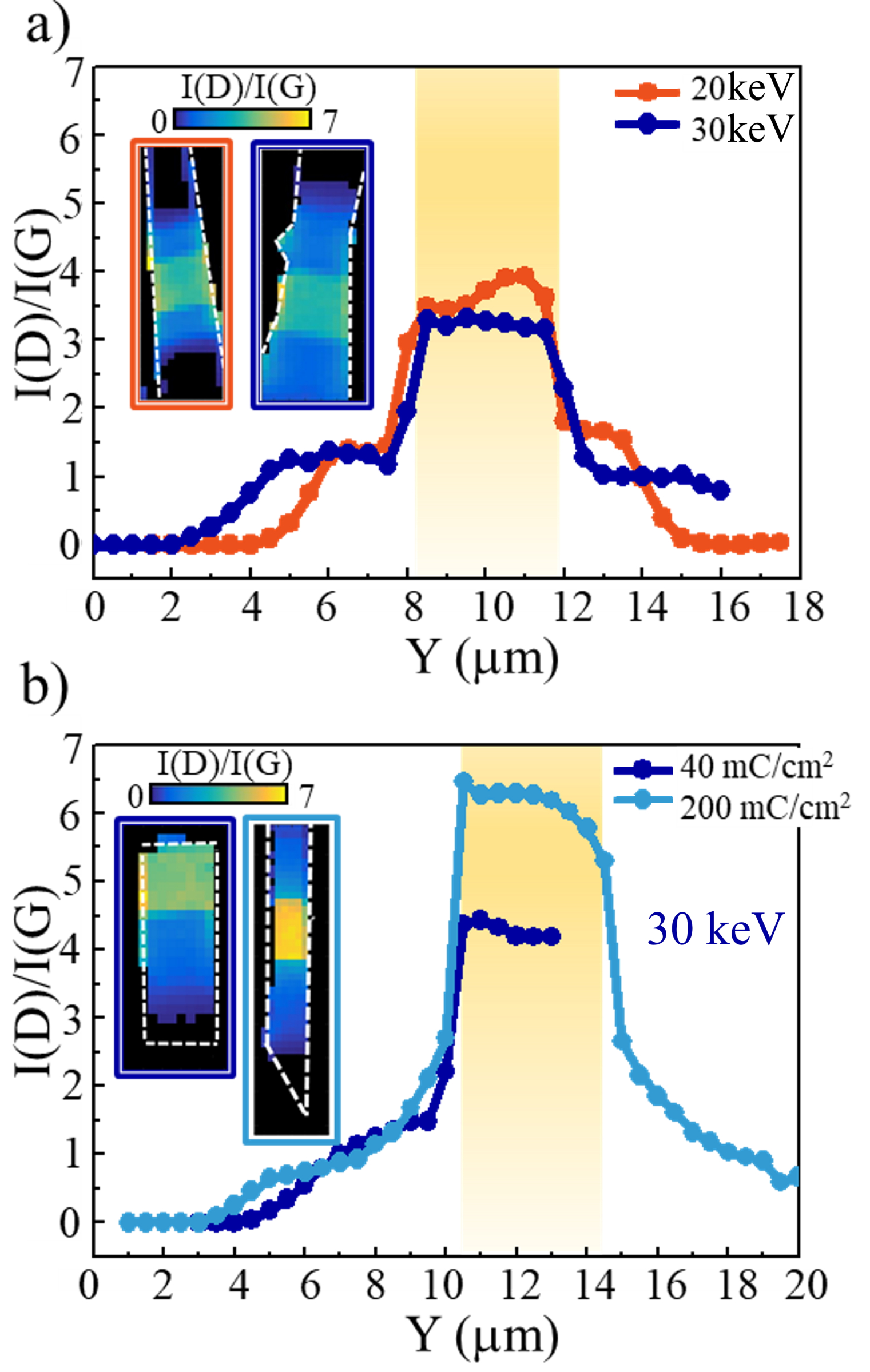}
  \caption{Unintentional defects dependency on irradiation conditions for graphene on no-plasma-treated substrate: a) averaged I(D)/I(G) profile along Y-axis under 20 keV (orange curve) and 30 keV (blu curve) electrons-irradiation. b) Averaged I(D)/I(G) profile along Y-axis as a function of the irradiation dose for electrons of 30 keV. Blu curve is for 40 mC/cm$^2$ and cyan curve is for 200 mC/cm$^2$.  In both panels, insets show the spatial distribution of I(D)/I(G) across the graphene flakes. The white dashed lines indicate the flake edges. The yellow area indicates the irradiated area.}
  \label{fig2}
\end{figure}

\begin{figure*}
\centering
  \includegraphics[height=7cm]{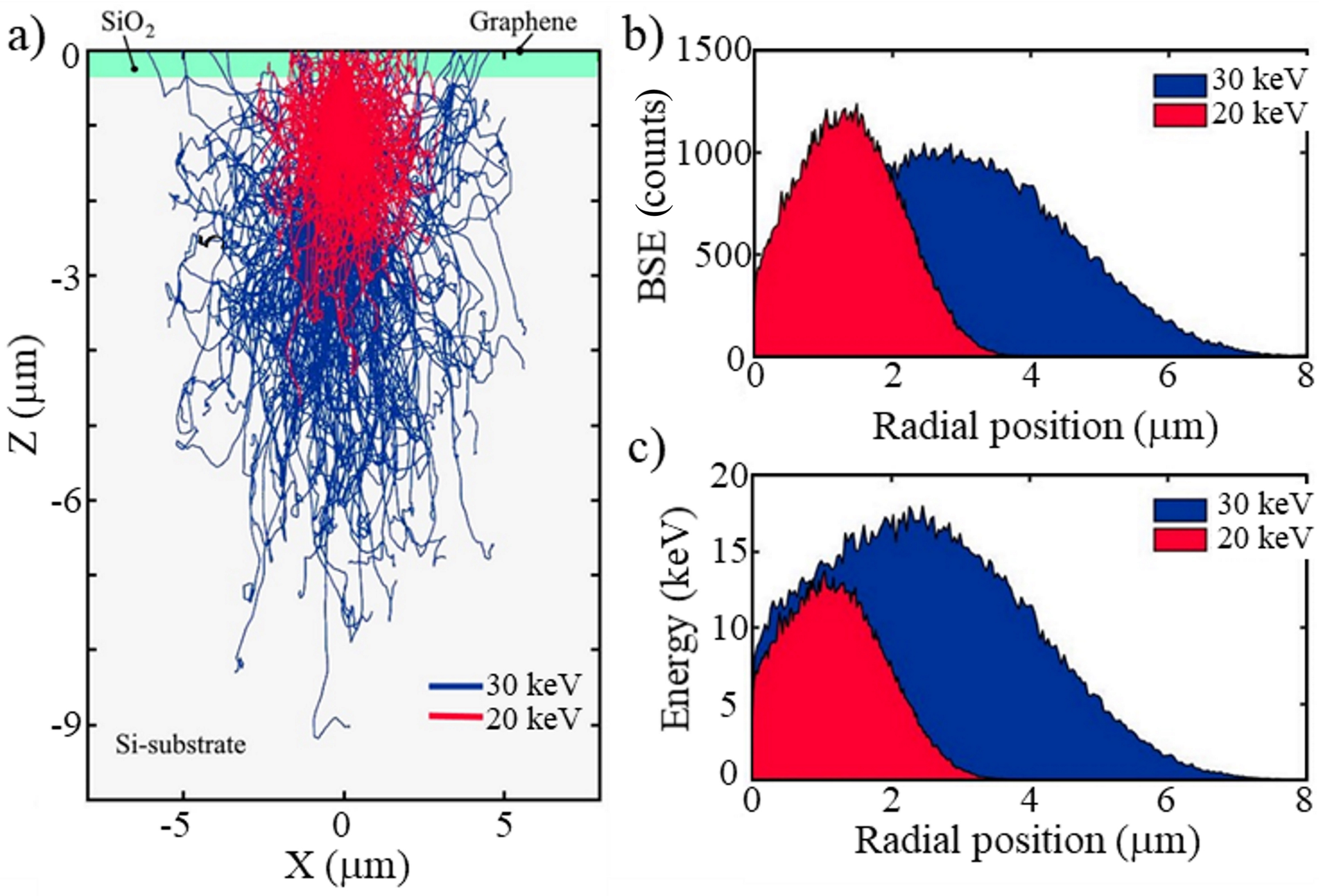}
  \caption{Monte Carlo simulations: a) representative 500 trajectories of primary electrons simulated considering 8$\times$10$^5$ electrons having kinetic energy of 20 keV (red curves) and 30 keV (blu curves). b) The number of escaping BSEs and c) the energy of BSEs as a function of the radial distance from the primary beam landing position simulated for kinetic energy of 20 keV (filled red curve) and 30 keV (filled blu curve).}
  \label{fig3a}
\end{figure*}
\section{Results and discussion}
Before electrons irradiation, the initial conditions of as-exfoliated graphene resulted in negligible intrinsic defects density (absence of Raman mode D, see black line in Fig. \ref{fig1}.a). When irradiating a portion of graphene sheet with electrons of 30 keV (violet area in the inset of Fig. \ref{fig1}.a), the extra D and D$^\prime$ peaks clearly appear in the Raman spectrum (violet line in Fig. \ref{fig1}.a), unambiguously indicating the lattice symmetry breaking. These bands are Raman active only when defects guarantee the momentum conservation and they are typically used to analyze the graphene disorder. In particular, the ratio between the intensity of D and G peaks [I(D)/I(G)] quantifies the defect density in crystalline-to-nanocrystalline regime \cite{Ferrari2007a, Bruna2014, Beams2015}. 

Figure \ref{fig1}.b shows the spatial distribution of I(D)/I(G) estimated across the graphene onto no-plasma-treated substrate under irradiation of electrons of 30 keV and dose $\sim$ 40 mC/cm$^2$. As expected, I(D)/I(G) reaches the maximum value  ($\sim$ 3.5) in the region exposed to the electron beam (zone 3), while it drops to zero in the unexposed area (zone 1). Assuming the model proposed in Ref. \cite{Bruna2014}, the measured I(D)/I(G) value corresponds to a density of defects of about 7$\times$10$^{11}$ cm$^{-2}$. Interestingly, the transition between zone 1 and zone 3 is not abrupt, but a region containing non-zero I(D)/I(G) values exists in between (zone 2). Considering the averaged profile along Y-axis (Fig. \ref{fig1}.c), I(D)/I(G) assumes a constant value ($\sim$ 1.5) in the first 2 $\mu$m far from the edge of the irradiated area and then monotonically decreases to zero. This transition area has an overall extension of about 5 $\mu$m and is homogeneously distributed along X-axis (see inset in Fig. \ref{fig1}.c). 

%\begin{figure*}
 %\centering
 %\includegraphics[height=3cm]{example2}
% \caption{A two-column figure.}
% \label{fgr:example2col}
%\end{figure*}

Graphene irradiated at 20 keV with the same dose shows similar behaviour: the creation of an area with monotonic decrease to zero of I(D)/I(G) that is coupled to an almost constant I(D)/I(G) zone (Fig. \ref{fig2}.a). Unlike 30 keV, irradiation by 20 keV-electrons generates a transition area having much smaller extension (about 2.5 $\mu$m), while the irradiated zones have comparable I(D)/I(G) values. A different evolution was observed when exposing graphene to 200 mC/cm$^2$ dose at 30 keV, as shown in Fig. \ref{fig2}.b. As expected, a much higher I(D)/I(G) value was quantified in the irradiated area due to the larger e-beam dwell time (140 ms compared to 30 ms for 40 mC/cm$^2$), while the transition zone expands for a longer distance (about 6-7 $\mu$m) compared to the smaller dose. It is worth to notice that the data for the 40 mC/cm$^2$ exposure in Fig. \ref{fig2}.b were acquired in a different flake compared to the one of Fig. \ref{fig1} and Fig. \ref{fig2}.a. This confirms the systematic creation of similar unintentional defects in the adjacency of irradiated areas.

Although electrons-irradiated graphene has been largely studied  \cite{Childres2010,Liu2011,Childres2014,Tao2013}, the analysis of the crystalline structure of the overexposed graphene edges has been only poorly investigated by two-dimensional mapping \cite{Gardener_2012}. In order to understand the origin of the observed extra defects, we simulated the scattering events of primary electrons within our structure for the two used kinetic energies, as shown by the trajectories reported in Fig. \ref{fig3a}.a. When multiple scattering events occur, some electrons can generate secondary electrons (SEs energy $<$ 50 eV) and some others can be scattered back towards the surface and escape at a distance that depends on the energy of the primary electrons (with energy $\sim$ 60-80$\%$ of primaries energy). As a result, BSEs are laterally spread with a radial distance, i.e. the distance of the escaping position from the landing point of the primary electrons, that can be of the order of few microns (Fig. \ref{fig3a}.b). When irradiating with kinetic energy of 20 keV, Monte Carlo simulations show that BSEs can escape at distance of the order of $\sim$ 3 $\mu$m, while at 30 keV this distance increases up to $\sim$ 6 $\mu$m due to the larger penetration depth of the primary electrons (Fig. \ref{fig3a}.a). Moreover, kinetic energy of BSEs has a functional dependency on the radial distance. As shown in Fig. \ref{fig3a}.c, BSEs from 20 keV primary electrons have maximum kinetic energy of about 10 keV at a radial distance of about 1 $\mu$m. Instead, BSEs having a maximum of 15 keV can escape at about 2.5 $\mu$m far from the primary electron landing point at 30 keV. 

\begin{figure}[t]
\centering
\includegraphics[height=7cm]{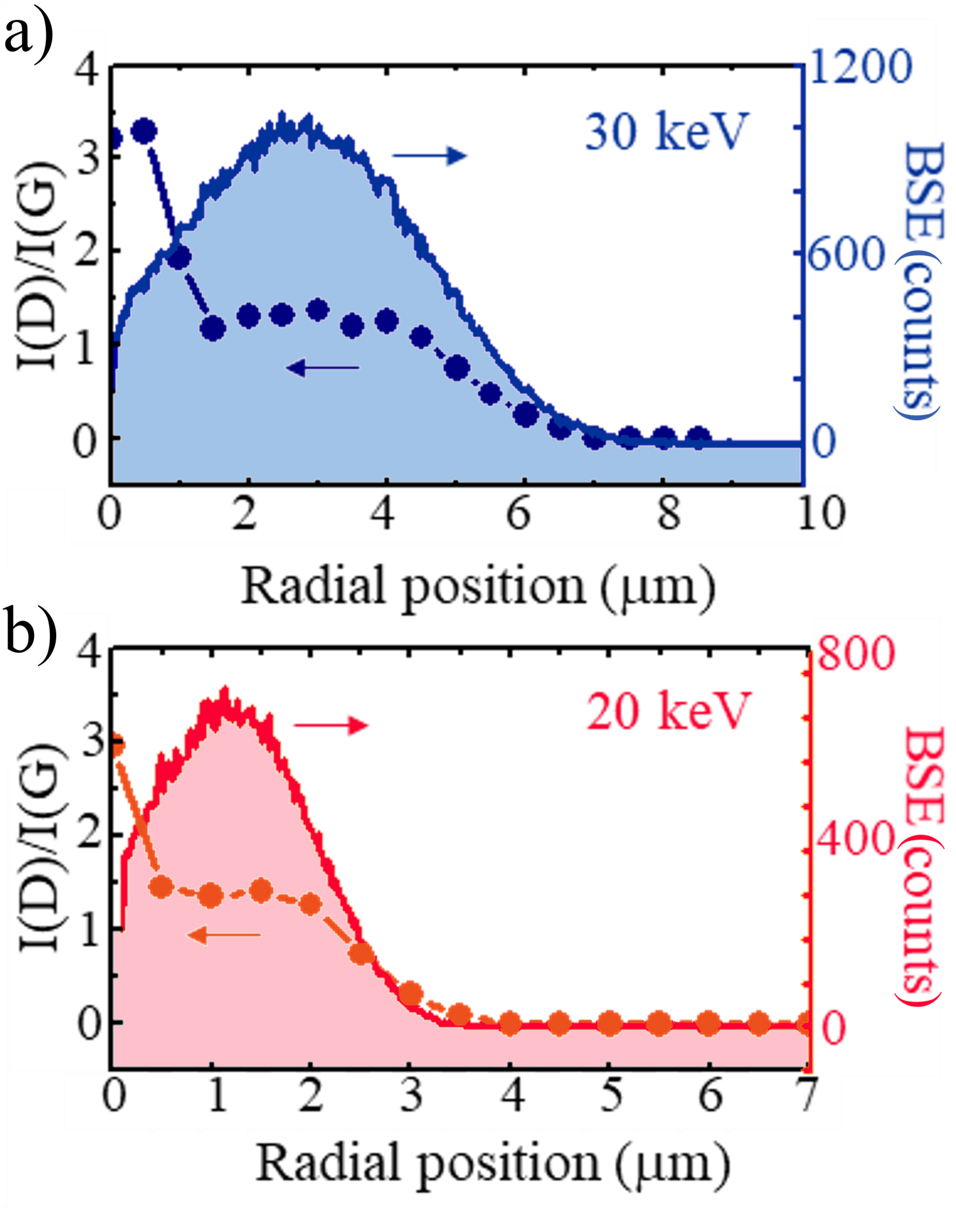}
\caption{Comparison between Monte Carlo simulations with experimental data: b) radial distance of escaping BSEs [panel a): filled blu curve for 30 keV; panel b): filled red curve for 20 keV] and spatial functional dependency of I(D)/I(G) values quantified starting from the edges of the irradiated area [panel a): dotted blu curve for 30 keV; panel b): dotted orange curve for 20 keV]}
\label{fig3b}
\end{figure}

By comparing the radial distribution of the number and kinetic energy of BSEs with the spatial dependency of I(D)/I(G) values within the transition area (zone 2), we found a very good agreement between the two distributions. As shown in Fig. \ref{fig3b}.a and Fig. \ref{fig3b}.b, for both primary electron kinetic energies the area where I(D)/I(G) has constant value corresponds to the region with the largest number of high energetic BSEs that escape from the surface. Indeed, irradiation in scanning mode implies a superposition of several BSEs radial distributions, thus an almost homogeneous dose of BSEs is expected within a radial distance of $\sim$1 $\mu$m and $\sim$3 $\mu$m at 20 keV and 30 keV, respectively. Moreover, the defective transition regions decay in good agreement with the spatial distribution of escaping BSEs.

%\subsection{Tables}
%Tables typeset in RSC house style do not include vertical lines. Table footnote symbols are lower-case italic letters and are typeset at the bottom of the table. Table captions do not end in a full point.\cite{Arduengo1992,Eisenstein2005}

\begin{figure}[h]
\centering
\includegraphics[height=10cm]{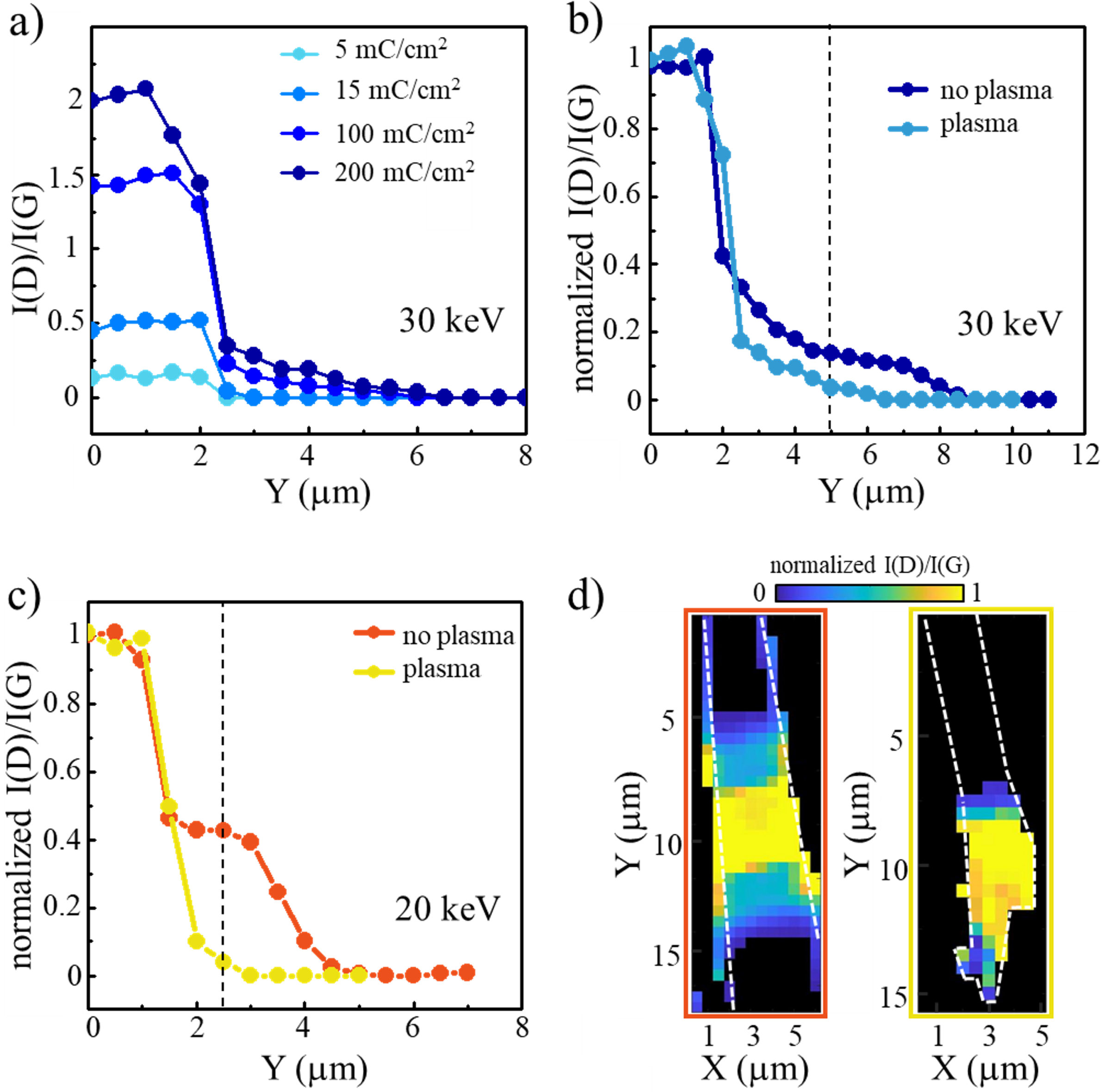}
\caption{Surface treatments effects: a) I(D)/I(G) profile along Y-axis for different doses (from 5 mC/cm$^{2}$ to 200 mC/cm$^{2}$) when irradiating graphene on plasma-treated substrate at 30 keV. b) Comparison between normalized I(D)/I(G) profiles of irradiated graphene on no-plasma-treated (blu dotted curve) and plasma-treated substrate (cyan dotted curve) for 30 keV and 200 mC/cm$^{2}$. c) Comparison between normalized I(D)/I(G) profiles of irradiated graphene on no-plasma-treated (orange dotted curve) and plasma-treated substrate (yellow dotted curve) for 20 keV and 40 mC/cm$^{2}$. The dashed-lines indicate the distance at which the largest number of BSEs escape from the surface according to Monte Carlo simulations. d) Two-dimensional spatial distribution of normalized I(D)/I(G) across the graphene flake on no-plasma- (left orange-framed panel) and plasma-treated (right yellow-framed panel) substrate, when irradiated at 20 keV with 40 mC/cm$^{2}$-electron dose. The white dashed lines indicate the flake edges.}
\label{fig4}
\end{figure}

Interestingly, similar experiments carried out on graphene exfoliated onto plasma-treated  substrates revealed very different results. As shown in Fig. \ref{fig4}.a, by increasing the exposure dose from 5 mC/cm$^{2}$ to 200 mC/cm$^{2}$ no strongly defective areas emerge outside the irradiated zones, except for the largest dose. This strong reduction of unintentional defective regions can be more clearly seen when comparing the normalized I(D)/I(G) trends for graphene on no-plasma- and plasma-treated substrates when exposing, for example, at 30 keV and 200 mC/cm$^{2}$ (Fig. \ref{fig4}.b) and at 20 keV and 40 mC/cm$^{2}$ (Fig. \ref{fig4}.c) or when plotting the two dimensional spatial distribution of the normalized I(D)/I(G) across the graphene flake. An example is given in Fig. \ref{fig4}.d when irradiating at 20 keV with 40 mC/cm$^{2}$. The data were normalized by the values of I(D)/I(G) in irradiated zones and summarized in Table \ref{tab1}. 

\begin{table*}
\small
  \caption{Maximum extension of unintentional detects zones for different primaries kinetic energies, electron doses and surface treatments.}
  \label{tab1}
  \begin{tabular*}{\textwidth}{@{\extracolsep{\fill}}lllllll}
    \hline
    \textbf{Defects extension} & \textbf{Kinetic energy} &  \textbf{Dose} &\textbf{Surface treatment} \\
\hline
3.5 $\mu$m &20 keV & 40 mC/cm$^2$ & no-plasma \\
1.5 $\mu$m & 20 keV & 40 mC/cm$^2$ & plasma\\
\hline
7 $\mu$m & 30 keV & 200 mC/cm$^2$ &  no-plasma\\
3.5 $\mu$m & 30 keV & 200 mC/cm$^2$ & plasma\\
    \hline
  \end{tabular*}
\end{table*}

The maximum extension of unintentional defects at 30 keV is about 3.5 $\mu$m in plasma-treated substrates and the maximum I(D)/I(G) of the transition zone is $\sim$0.15 times the one in the irradiated area, while it was 6-7 $\mu$m and a factor $\sim$0.3 in no-plasma-treated substrates, respectively. At 20 keV the maximum radial distance reachable in plasma-treated substrates is about 1.5 $\mu$m, which is 2 $\mu$m-shorter compared to no-plasma-treated substrates and the I(D)/I(G) value is a factor $\sim$0.1 of the irradiated zone, instead of $\sim$0.4 in no-plasma-treated substrates. Moreover, unlike in no-plasma-treated substrates, the non-zero I(D)/I(G) zones in plasma-treated substrates do not precisely reproduce the spatial distribution of arrival BSEs, but originate only within the distances where the largest number of BSEs escapes from the surface, as shown in Fig. \ref{fig4}.b and Fig. \ref{fig4}.c. Consequently, the BSEs appear having only a minor effect on the adjacent areas of the irradiation in plasma-treated substrates, suggesting the contribution from additional defects sources in no-plasma-treated substrates.

A possible explanation to the obtained data can involve secondary electrons. As already mentioned, secondary electrons can be generated during the scattering events of primary electrons passing through solids. However, the penetration depth of primary electrons in silicon is much larger than the escape depth of SEs, thus their contribution in forming SEs is expected to be low. Instead, the BSEs can dissipate their energy in SEs escape region, thus they extensively contribute to SEs creation even when the backscattering coefficient is relatively small \cite{Kanter1961}. Conventionally, SEs are defined as particles having energy lower than 50 eV. It is worth mentioning that this energy range matches with the energy peak of dissociation cross section of molecules like hydrocarbons or other organic molecules \cite{Kim2014, Kim2016}. For example, complete or partial hydrogenation process mediated by SE-induced fragmentation of adsorbed H$_2$O molecules has been reported in electron-irradiated graphene \cite{Jones2009, Jones2010}. Hence, under electron irradiation, adsorbants dissociation can occur onto solids surface, creating reactive radicals far from the landing point of primary electrons and within a spatial length determined by the escaping radial distance of BSEs. Consequently, when irradiating graphene placed onto surfaces (here SiO$_2$) with residual organic compounds, the SEs-formed radicals are trapped at the interface between graphene and SiO$_2$ and are expected to interact with the graphene lattice and/or its defects, promoting the formation of Raman-active bond deformations, and thus the appearance of D peak in the Raman spectrum, and charge transfer to graphene \cite{Kim2014, Wang2019}. Based on these considerations, the large D peak in the transition zones observed in no-plasma-treated substrates may be explained by the presence of SEs-dissociated organic contaminants that are trapped at the interface between graphene and SiO$_2$ \cite{Saga1997}. Indeed, while oxygen plasma cleaning produces SiO$_2$ surfaces poor of adsorbents due to the strong action of plasma activated species \cite{Shunko2007}, the solvent stripper, like specific resist remover, is less effective in removing most organic contaminants. 
\\To further corroborate our hypothesis, we also analyzed the doping condition of the transition zones, because, as mentioned, radicals can induce a charge doping in graphene sheet. An unambiguous Raman fingerprint of doping status in graphene is represented by the width of G peak [$\Gamma$(G)] \cite{Casiraghi2007}. In pristine graphene, $\Gamma$(G) value is mainly determined by graphene chemical potential, as the possible electron-hole pairs formation is controlled by Pauli blocking. In intrinsic graphene, $\Gamma$(G) $\sim$ 16 cm$^{-1}$, whereas $\Gamma$(G) symmetrically decreases as electrons or holes concentration increases due to the limited phonon decay paths \cite{Beams2015}. Instead, in electron-irradiated graphene,  $\Gamma$(G) value arises from the competitive action of  carrier concentration (G peak narrowing) and defects density (G peak broadening), where the latter mainly dominates for I(D)/I(G) $>$ $\sim$3 \cite{Childres2014}.

\begin{figure}[h]
\centering
\includegraphics[height=10cm]{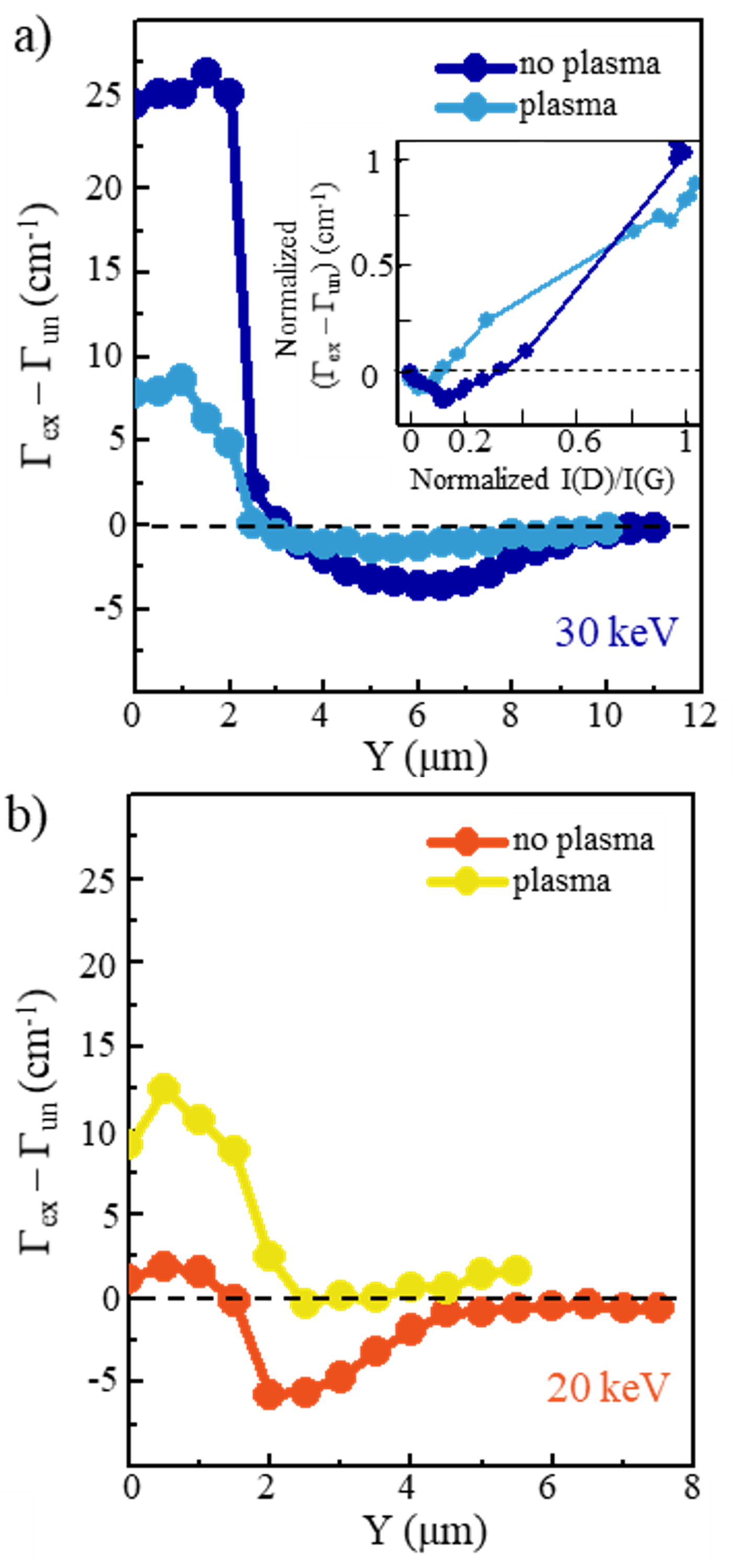}
\caption{Irradiation effects on G peak width: a) profile of the difference between G-peak width in 30 keV-exposed ($\Gamma_{ex}$) and as-exfoliated ($\Gamma_{un}$) graphene along Y-axis in no-plasma- (blue dotted-line) and plasma-treated (cyan dotted-line) substrates. The electron dose was 200 mC/cm$^2$. Inset: normalized $\Gamma_{ex}$ - $\Gamma_{un}$ as a function of normalized  I(D)/I(G).  b) Profile of the difference between G-peak width in 20 keV-exposed ($\Gamma_{ex}$) and as-exfoliated ($\Gamma_{un}$) graphene along Y-axis in no-plasma- (orange dotted-line) and plasma-treated (yellow dotted-line) substrates. The electron dose was 40 mC/cm$^2$.}
\label{fig5}
\end{figure}

Figure $\ref{fig5}$ shows the Y-axis profile of the difference of $\Gamma_G$ with respect to its value in as-exfoliated graphene in plasma- and no-plasma-treated substrates when irradiating with 30 keV (Fig. $\ref{fig5}$.a) and 20 keV (Fig. $\ref{fig5}$.b). For both electron kinetic energies, the irradiated areas have G peak width ($\Gamma_{ex}$) larger than in as-exfoliated ($\Gamma_{un}$) graphene both in plasma- and no-plasma-treated substrates, as expected due to the larger defects-dominated scattering contribution \cite{Childres2014}. Instead, a narrowing of G peak is observed in the transition zones of graphene onto no-plasma-treated substrates, confirming the doping of graphene in these areas. This result is also supported by the analysis of G peak width as a function of I(D)/I(G). As shown in the inset of Fig. $\ref{fig5}$.a, G peak width firstly decreases for low defect density (corresponding to the transition area) and then increases reaching its maximum value in the irradiated area. The data of $\Gamma_{ex} - \Gamma_{un}$ and I(D)/I(G) in inset of Fig. $\ref{fig5}$.a were normalized by their maximum values in the irradiated zone. Similar data are obtained also at 20 keV. Additionally, these narrower $\Gamma_G$ values have Y-axis profiles that follow the spatial distribution of BSEs, like for I(D)/I(G), reaching the minimum value at the same radial distance at which the maximum number of BSEs escapes from the surface. Instead, in plasma-treated-substrates $\Gamma_G$ is almost equal to the value of as-exfoliated graphene, suggesting a negligible doping related to the electrons irradiation. In both substrates, G peak has almost unaltered width when considering zones very far from the patterned areas, indicating a conservation of the graphene conditions during the entire experiments. Hence, the combination of I(D)/I(G) and $\Gamma_G$ data pointed out the relevant role of BSEs and SEs in affecting the lateral resolution of defects-patterning in graphene sheets: BSEs possess sufficient kinetic energy to introduce defects in graphene lattice, as revealed in plasma-treated substrates; BSEs-generated SEs, instead, induce molecular dissociation of surface organic residues, that yields to both structural defects and charge doping, as clearly observed in no-plasma-treated substrates.

\section{Conclusions}
Here, we created patterned structural defects on graphene sheets by electron-beam irradiation and analysed their distribution by spatially-resolved micro-Raman spectroscopy. Surface treatments of the graphene-supporting substrate have strong impact on the lateral resolution that can be achieved on the final defective pattern. Micrometer-large-unintentional defects-rich zones were revealed in the adjacent parts of the irradiated areas and have I(D)/I(G) spatial distribution that strongly depends on primary electron dose and kinetic energy, but also on SiO$_2$ surface cleaning methods. Indeed, in no-plasma-treated substrates, the irradiation-surrounding areas have very large I(D)/I(G) values with a few-$\mu$m-long lateral extension, whereas plasma-treated substrates areas show smaller I(D)/I(G) values in a shorter distance from the primaries impact point. Simulation of primary electrons scattering events via Monte Carlo method demonstrated that these transition zones originate within the area where BSEs and BSEs-created SEs escape from the SiO$_2$ surface. In no-plasma-treated substrates, the radial distribution of unintentional defective areas is in very good agreement with the simulated one of BSEs. In these substrates, the I(D)/I(G) values is larger and has longer extension due to the combined action of BSEs and SEs. BSEs have sufficient energy to still induce defects on the graphene lattice, instead SEs dissociate the possible organic residues trapped between graphene and SiO$_2$ surface. These SEs-created radicals, interacting with the graphene sheet, cause additional structural defects and charge doping. The latter was confirmed by analysing the G peak width. Instead, in plasma-treated substrates, only BSEs affect the lateral resolution of defects-patterning, as the plasma treatment removes the impurities adsorbed on the substrate. In general, BSEs contribution can be mitigated by using very low Z-number materials or, alternatively, by reducing the thickness of the substrates. 

\section*{Conflicts of interest}
There are no conflicts to declare.

\section*{Acknowledgements}
The authors wish to acknowledge Dr. S. Heun from Istituto Nanoscienze-Consiglio Nazionale delle Ricerche for the fruitful discussion.
The authors wish to acknowledge Dr. C. Coletti from Istituto Italiano di Tecnologia for the access to micro-Raman system.

%%%REFERENCES%%%

\end{document}